\documentclass[fleqn,twoside]{article}
\usepackage{espcrc2}

\usepackage[figuresright]{rotating}

\RequirePackage{xspace}

\newcommand{\AmS}{{\protect\the\textfont2
  A\kern-.1667em\lower.5ex\hbox{M}\kern-.125emS}}
\newcommand{\Dbar}{\kern 0.18em\overline{\kern -0.18em D}{}\xspace}
\newcommand{\DD}{\ensuremath{D\Dbar}\xspace}
\newcommand{\Dz}{\ensuremath{D^0}\xspace}
\newcommand{\Dzb}{\ensuremath{\Dbar^0}\xspace}
\newcommand{\DzDzb}{\ensuremath{\Dz {\kern -0.16em \Dzb}}\xspace}
\newcommand{\DsP}{\ensuremath{D_S^+}\xspace}
\newcommand{\DsM}{\ensuremath{D_S^-}\xspace}
\newcommand{\DspDsm}{\ensuremath{\DsP {\kern -0.16em \DsM}}\xspace}
\newcommand{\Dp}{\ensuremath{D^+}\xspace}
\newcommand{\Dm}{\ensuremath{D^-}\xspace}
\newcommand{\DpDm}{\ensuremath{\Dp {\kern -0.16em \Dm}}\xspace}

\title{Charm Physics with BES-III at BEPC-II}
\author{Haibo Li\address{Institute of High Energy Physics,
           P.O.Box 918, Beijing  100049, China} \thanks{Email:
lihb@ihep.ac.cn}}
\begin{document}
\begin{abstract}
We report on the charm physics potential at BES-III at BEPC-II which
will make significant contribution to quark flavor physics this decade. 
The charm physics program will include studies of leptonic, semileptonic and
hadronic charm decays, and tests for physics beyond the Standard Model.
Event samples of the order of 30 million $\DD$
pairs, 2 million $\DspDsm$ pairs at threshold and $10\times 10^9$ $J/\psi$
decays will be produced with one year design
luminosity. High precision charm data will enable us to validate
forthcoming Lattice QCD calculations at the few percent level. 
These can then be used to make precise measurements of CKM elements,
$V_{cd}$, $V_{cs}$, $V_{ub}$, $V_{cb}$ and $V_{ts}$, which are useful to improve the accuracy of test of 
the CKM unitarity.  
\vspace{1pc}
\end{abstract}

\maketitle

\section{Introduction}

BES-III at BEPC-II under construction in Beijing 
will focus on charm and QCD physics in the range $\sqrt{s} = 2.0 - 4.2$ GeV. 
The physics program at BES-III includes a variety of measurements that
will improve the understanding of the Standard Model (SM) processes as well
as provide the opportunity to probe for physics that lies beyond the
SM. However, the possibility of new physics may
be diluted by the large hadronic uncertainty in $D$ decays. The study of weak
interactions in the charm sector, and the extraction of quark mixing matrix parameters remain
limited by our capacity to cope with non-perturbative QCD. Thus, it will be
of great benefit if we can understand the effects of strong
coupling in QCD. High precision predictions of QCD will then remove road
blocks for many weak and flavor physics measurements. 

Lattice QCD (LQCD)~\cite{millc} had matured over the last decade. Recent
advances in LQCD have produced a wide variety of calculations of
non-perturbative quantities with accuracies
in the 10-20\% level for $D$ decay constant and form factors.
We finally can expect the first LQCD results with 1-3\% errors. BES-III
will provide crucial data to validate them and help guide the development of 
QCD calculation techniques for a full understanding of non-perturbative QCD effects. This will
improve the measurements of charm meson decay constants, $V_{cs}$ and
$V_{cd}$ to 1\% level.
\begin{table}[htbp]
\caption{$\tau$-Charm productions at BEPC-II in one year's running ($10^7s$).} 
\label{tab:lum}
\begin{tabular}{@{}lll}
\hline
               & Central-of-Mass    & \#Events  \\
Data Sample    & (MeV)                   & per year  \\
\hline
$J/\psi$ &  3097     & $10\times 10^9$\\
$\tau^+\tau^-$   & 3670  & $12\times 10^6$ \\
$\psi(2S)$ & 3686  & $3.0\times 10^9$ \\
$\DzDzb$ & 3770  & $18\times 10^6$ \\
$\DpDm$ & 3770  & $14\times 10^6$ \\
$\DspDsm$ & 4030  & $1.0\times 10^6$ \\
$\DspDsm$ & 4170  & $2.0\times 10^6$ \\
\hline
\end{tabular}
\end{table}

Beginning in mid-2007, the BEPC-II accelerator will be
operated at center-of-mass (CM) energies corresponding
to $\sqrt{s} = 2.0 -4.2 $ GeV. The designed luminosity over this energy region will
range from $1\times 10^{33}$cm$^{-2}$s$^{-1}$ down to about $0.6 
\times 10^{33}$cm$^{-2}$s$^{-1}$~\cite{bepcii}, yielding around 5 fb$^{-1}$ each at 
$\psi(3770)$ and at $\sqrt{s} = 4170$ MeV~\cite{new_scan} above $\DspDsm$ threshold and
3 fb$^{-1}$ at $J/\psi$ peak in one year's running with
full luminosity~\cite{bepcii}. These integrated
luminosities correspond to samples of 2.0 million $\DspDsm$,
30 million $\DD$ pairs and $10\times 10^9$ $J/\psi$ decays. Table~\ref{tab:lum}
summarizes the data set per year at BES-III. In
this report, all preliminary sensitivity studies are based
on 20 fb$^{-1}$ luminosity at $\psi(3770)$ peak for $D$ physics, the same
luminosity also for $D_S$ physics at  $\sqrt{s} = 4170$ MeV.

\section{Unique Features of the Charm Physics at BES-III}

Many of the measurements related to charm decays have been done by other
experiments such as BES-II and CLEO-c, and many are also accessible to the
B-factory experiments. What are BES-III's advantages to running at the
open charm threshold? 

Firstly, compared to the old BES experiment, BES-III
is a state-of-the-art detector with modern technologies~\cite{bepcii}.
BES-I/BES-II was modeled
on MARK-III detector. It has gone through several generations
of development since then. In every resolution and performance
parameter, for example, hit resolution, momentum and energy resolutions,
mass resolution, particle identification (PID) capability, solid angle coverage, BES-III
is superior to previous versions of BES detectors by substantial margins. 
The new detector consists of a He-based small cell drift
chamber, Time-Of-Flight (TOF) counters for
PID, a CsI(Tl) crystal calorimeter, a solenoid
super-conducting magnet with a field of 1 Tesla and the magnet
yoke interleaved with RPC counters as the muon chamber. The construction
is expected to be completed in the middle of 2007.
Photon energy resolution is $\Delta E/E = 2.5\%$ at $E_{\gamma}= 1.0$ GeV. 
The momentum resolution is $\sigma_p/p = 0.5\%$ at $p= 1.0$ 
GeV/$c$, and the $dE/dx$ resolution for hadron tracks is about 6\%. 
The time resolution of TOF is about 100 ps,  combining energy loss ($dE/dx$) measurement in the draft
chamber, give 10 $\sigma$ K/$\pi$ resolution across the typical kinematic range.
A 25\% increase of solid angle coverage relative to BES improve efficiency
for double-tag measurements greatly.  The gains go as
$1.25^N$, where $N$ is the total number of tracks and photons
of the event~\cite{gibbons}.
This indicates a typical effective luminosity gain of 8 in such analyses.
For partial wave analysis, the increase in solid angle
coverage means angular distributions will be measured across the
almost full angular range without large variations
in acceptance. The $J^{PC}$ and partial waves can be measured
reliably and precisely.  

Secondly, BES-III will not be able to compete both BaBar and Belle
in statistics on charm physics, especially on the rare and forbidden decays
of charm mesons. However, data taken at charm threshold still have powerful
advantages over the data at $\Upsilon(4S)$, which we list
here~\cite{gibbons}: 
1) Charm events produced at threshold are extremely clean;
2) The measurements of absolute branching fraction can be made by using 
   double tag events;
3) Signal/Background is optimum at threshold; 
4) Neutrino reconstruction is clean;
5) Quantum coherence allow simple~\cite{gronau} and complex~\cite{asner} methods to measure $\DzDzb$ mixing
parameters and check for direct $CP$ violation. 

\section{Charm Decays and Production Cross Sections at BEPC-II}
\label{sec:Charm_decay}

The main targets of the charm physics program at BES-III are
absolute branching fraction measurements of leptonic, semileptonic and
hadronic decays. The first measures decay constants and the second measures
 form factors and, in combination with theory, allows the determination
of $V_{cs}$ and $V_{cd}$. The third of those provides an
absolute scale for all charm and hence beauty decays.

At $\DD$ or $\DspDsm$ threshold, no additional hadrons accompanying
 the $\DD$ or $\DspDsm$ pairs are produced. Reconstruction of one $D$ or
 $\Dbar$ meson (called single tag or ST) tags the event as either $\DzDzb$
or $\DpDm$ ($\DspDsm$). For a given decay mode $i$,
we measure independently the $D$ and $\Dbar$ ST yields, denoted
by $N_i$ and $\overline{N}_i$. We determine the corresponding
efficiencies from Monte Carlo simulations (MC), denoted by
$\epsilon_i$ and $\overline{\epsilon}_i$. Thus, $N_i = \epsilon_i {\cal
B}_i N_{\DD}$ and $\overline{N}_i = \overline{\epsilon}_i {\cal B}_i N_{\DD}$, 
where ${\cal B}_i$ is the branching fraction for mode $i$, assuming no
$CP$ violation, and $N_{\DD}$ is the total number of produced $\DD$ pairs 
 at BES-III. Double tag (DT) events are the subset of ST events where both
the $D$ and $\Dbar$ are reconstructed. The DT yield for $D$ mode $i$ and
$\Dbar$ mode $j$, denoted by $N_{ij}$, is given by $N_{ij} = \epsilon_{ij}
{\cal B}_i {\cal B}_j N_{\DD}$, where $\epsilon_{ij}$ is the DT efficiency. As
with ST yields, the charge conjugate DT yields and efficiencies, $N_{ji}$ and
$\epsilon_{ji}$, are determined separately.
Charge conjugate particles are implied, unless referring to ST and
DT yields.  The absolute branching fraction ${\cal B}_i$ can be obtained from
$\displaystyle {\cal B}_i = \frac{N_{ij}}{\overline{N}_j}
\frac{\overline{\epsilon}_j}{\epsilon_{ij}}$.
With the same method, we have the total number of $\DD$ pairs 
$\displaystyle N_{\DD} =\frac{N_i \overline{N}_j}{N_{ij}}\frac{\epsilon_{ij}}{(\epsilon_i
\overline{\epsilon}_j)}$, which can be used to obtain the absolute
cross-section of $\DD$ production.

\subsection{Leptonic Charm Decays}

From the leptonic decays
of $D^\pm$ and $D^{\pm}_S$ mesons, the decay constants $f_D$ and $f_{D_S}$
can be determined to a precision of about 1\%. The decay constants measure 
the non-perturbative wave function of the meson at
$zero$ inter-quark separation and appear in all processes where constituent
quarks must approach each other at distances small compared to the meson
size~\cite{gibbons}. 
\begin{table}[htbp]
\caption{Expected errors on the branching fractions for leptonic
decays and decays constants at BES-III with 20 fb$^{-1}$ at $\psi(3770)$ peak.}  
\label{tab:constants}
\begin{tabular}{@{}lll}
\hline
               & Error (\%)      & Error (\%)  \\
Decay Modes     &  on ${\cal B}$ & on $f_{D(s)}$  \\
\hline
$D^+ \rightarrow \mu^+ \nu$ ($f_D$)      &  2.0     & 1.5     \\
$D^+_S \rightarrow \mu^+ \nu$ ($f_{D_S}$)  &  2.0 & 1.1 \\
$D^+_S \rightarrow \tau^+ \nu$ ($f_{D_S}$) & 1.5  & 0.9 \\
\hline
\end{tabular}
\end{table}

Measurements of leptonic decays 
at BES-III will benefit from the fully tagged $D^+$ and $D^+_S$ decays
available at the $\psi(3770)$ and at $\sqrt{s} \sim 4170$
MeV~\cite{new_scan}. The leptonic decay of $D^+ (D^+_S) \rightarrow \mu^+
\nu$ is detected in tagged events by observing a single charged track
of the correct sign, missing energy, and a complete accounting
of the residual energy in the calorimeter. The pure $\DD$ pair
in the initial state and cleanliness of the full tag reconstruction make
this measurement essentially background-free.  The
leptonic decay rates for $D^+$ and $D^+_S$ can be measured
with a precision of 1-2\% level. This will allow the validation of
theoretical calculations of the decay constants at the 1\%
level. Table~\ref{tab:constants} summarizes the expected precision
in the decay constant measurements.  It should be noted that the
$D^+ \rightarrow \tau^+ \nu$ decay is reported by CLEO-c with upper limit 
of $2.1 \times 10^{-3}$ at 90\% CL~\cite{taunu}. At BES-III, the sensitivity will
be $10^{-5} - 10^{-6}$ level. 

\subsection{Semileptonic Charm Decays}

Semileptonic widths for $D \rightarrow X_{s(d)} l^+ \nu$
directly probe the elements of the CKM matrix.
When $J^P(X_{s,d}) = 0^-$, the differential width is given by:
\begin{equation}
\frac{d\Gamma(D \rightarrow X_{s(d)} l^+ \nu)}{dq^2} =
\frac{G^2_F}{24\pi^3}p^3 |V_{cs(d)}|^2 p^3 |F(q^2)|^2, 
\label{eq:semi}
\end{equation}
where $q^2$ is the momentum transfer squared between the $D$ meson and the
final state hadron in the $D$ rest frame, and $F(q^2)$ is the hadronic
form factor at $c \rightarrow W s(d)$ vertex. The form factor can be
predicted from a number of different theoretical
approaches~\cite{isgur} including LQCD~\cite{millc}. In addition
to its own intrinsic interest, the analogous form factor is needed for
extracting $b$-quark matrix elements such as $V_{ub}$ in $B$
semileptonic decays, so it is important to demonstrate the reliability
of any one calculation. 
\begin{table*}[htb]
\caption{Uncertainties
on the branching fractions for $D$ and $D_S$
semileptonic decay modes and precision of form
factor parameters at BES-III (assuming 20 fb$^{-1}$ data at
$\psi(3770)$ peak). The precision of parameters is mainly limited
by the uncertainties on $V_{cd}$ and $V_{cs}$.}
\label{tab:semi_decays}
\begin{tabular}{@{}lllll}
\hline
                &  Error (\%)   & Expected Error (\%)  &  Form Factor  & Expected Error (\%)     \\
Decay Modes     & on ${\cal B}$ PDG2004 & on ${\cal B}$ at BES-III & Type  & on Form
factor at BES-III \\
\hline
$D^0 \rightarrow K l\nu$  & 4.6 & 0.2 & PS$\rightarrow$ PS &  1.0 \\
$D^0 \rightarrow \pi l \nu$ & 9.6 & 0.4 &PS$\rightarrow$ PS & 1.0 \\
$D^+ \rightarrow \pi l \nu$  & 50.0 & 0.8 &PS$\rightarrow$ PS & 1.0  \\
$D^+ \rightarrow \overline{K}^* l \nu$  & 10.0 & 0.3 & PS$\rightarrow$ V & 2.0 \\
$D^+_S \rightarrow \phi l \nu $   & 25.0& 1.2 & PS$\rightarrow$ V & 1.0\\
\hline
\end{tabular}
\end{table*}

Absolute branching ratios in critically interesting modes,  such as,
$D \rightarrow \pi l\nu$, $D \rightarrow K l\nu$, $D \rightarrow
\eta (\eta^\prime) l\nu$,  
$D \rightarrow \rho l \nu$,
$D \rightarrow K^* l \nu$, $D_S \rightarrow \phi l\nu$ and $D_S \rightarrow K^* l\nu$, 
will be measured to be less than 1\%, and the form
factor slopes to 1.5\%. The measurement in each case is based on the use
of tagged events where the cleanliness of the environment provides nearly
background-free signal samples. 
This will lead to the determination of the CKM matrix elements $V_{cs}$ 
and $V_{cd}$ with a precision of 1\% level assuming knowledge of the relevant 
form factors with 1.5\% uncertainties from LQCD.
Form factors in all modes can be measured across the full range of $q^2$ with excellent resolution. 
Measurements of the vector and axial-vector form factors $V(q^2)$,
$A_1(q^2)$ and $A_2(q^2)$ will also be possible at the $\sim 2\%$ level.
Table~\ref{tab:semi_decays} summarizes the most recent results and the expected fractional errors
on the branching ratios at BES-III. 

With high statistics of $D$ sample at BES-III, many unobserved
semileptonic $D$ decays will be accessible, such
as $D+ \rightarrow f_0 l\nu$, $D+ \rightarrow \eta
(\eta^{\prime}) l \nu$, $D^+ \rightarrow \overline{K}^{**} l \nu$ and $D^+ \rightarrow
\phi l\nu$, where $\overline{K}^{**}$ is the exited koan
mesons including $\overline{K}_1(1270)$,
$\overline{K}^{*}(1410)$ and $\overline{K}^{*}(1430)$.  These measurements
will provide another lab to study the $K\pi$ and $\pi\pi$ S-wave~\cite{yang}. 

The ratio of decay widths
$\Gamma(D_S \rightarrow \eta^{\prime} e \nu)/\Gamma(D_S \rightarrow
\eta e \nu)$, using $|V_{cs}| = 0.975$~\cite{bugg},
\begin{equation}
\frac{\Gamma(D_S \rightarrow \eta^{\prime} e \nu)}{\Gamma(D_S \rightarrow
\eta e \nu)} \sim 0.28 \times |cot\phi|^2, 
\label{eq:eta-etap}
\end{equation}
depends on the content of the $\eta$ and $\eta^\prime$ mesons, where $\phi$ is the mixing angle in the $\eta - \eta^{\prime}$ flavor
mixing scheme~\cite{feldman}. We note that the decays $D_S \rightarrow 
 \eta(\eta^\prime) l \nu$ involve the strange content of $\eta(\eta^\prime)$,
and  $D^+ \rightarrow  \eta(\eta^\prime) l \nu$ involve the non-strange
content. Therefore,  $D^+ \rightarrow
 \eta(\eta^{\prime}) l \nu$ and $D_S \rightarrow \eta(\eta^{\prime}) l \nu$ 
could provide combined testing of a $\eta -\eta^{\prime}$ mixing
scheme~\cite{yyd}. The uncertainty on mixing angle $\phi$ is about 13\%
with current $D^+_S$ data~\cite{feldman}, at BES-III, the error will be 2\% level. 

\subsection{Absolute Hadronic Branching Fraction and $e^+e^- \rightarrow
\DD$,$D\overline{D}^*$, and $D^*\overline{D}^*$ Cross Sections}

Absolute branching fraction measurements are important since, for a lot of analyses
at higher energies as well as in the $B$-system, an
inaccurate knowledge of $D$, $D_S$ decays can result in large
systematic errors. Using double-tagged events at threshold leaves
only major systematic error contributions from efficiency uncertainties in
the tracks and showers.
\begin{table}[htbp]
\caption{Unobserved 3-body $D$ decays,
$D^0 \rightarrow P^0 P^0 X^0$, where $P^0$ is pseudoscalar and $ X^0$
is any kind of particle allowed in the final states.
$f_0$ and $a_0$ are the scalar $f_0(980)$ and $a_0(980)$, respectively.
N/A represents not available.} 
\label{tab:miss_3}
\newcommand{\cc}[1]{\multicolumn{4}{c}{#1}}
\begin{tabular}{@{}c|c|c|c|c}
\hline
             &  \cc{$P^0$} \\
\hline
 $X^0$       &  $\pi^0$ & $\eta$ & $K_S$  & $K_L$    \\
\hline
$\pi^0$ &   $\pi^0 \pi^0 \pi^0$ & $\eta \eta \pi^0$ & $K_S K_S \pi^0$ & $K_L K_L \pi^0$ \\ \hline
$\eta$  & $\pi^0 \pi^0 \eta$ & $\eta \eta \eta$ & $K_S K_S \eta$ & $K_L K_L \eta$ \\ \hline
$\eta^\prime$ & $\pi^0 \pi^0 \eta^\prime$ & N/A & N/A & N/A \\\hline
$K_S$   & $\pi^0 \pi^0 K_S$ &  $\eta \eta K_S$ &  $K_S K_S K_S$ & $K_L K_L K_S$ \\\hline
$K_L$ &  $\pi^0 \pi^0 K_L$ & $\eta \eta K_L$ & $K_S K_S K_L$ & $K_L K_L K_L$ \\\hline
$a_0$ &  $\pi^0 \pi^0 a_0$ &  N/A & N/A & N/A   \\\hline
$f_0$ &  $\pi^0 \pi^0 f_0$ &  N/A & N/A & N/A   \\
\hline
\end{tabular}
\end{table}

The rate for the critical normalizing modes $D \rightarrow K\pi$, $D^+
\rightarrow K\pi\pi$, and $D_S \rightarrow \phi \pi$ will be established
to a precision of order less than 1.0\%.  At BES-III, the statistic is high
enough that we can measure Cabibbo-suppressed decays
of $D$ mesons, especially the $D \rightarrow 4 \pi$, $5 \pi$ or even
more pions final states~\cite{kspipi}. The sensitivities of the measurements
of these branching fractions will be $10^{-5} - 10^{-6}$ level at BES-III.  
Many unobserved 3-body $D$ decays can also be observed at BES-III as listed
in Table~\ref{tab:miss_3}. One should note that the decays
$D (D_S) \rightarrow V_1 V_2$ are very important
to probe the final state interaction by measuring the polarization
fractions, $f_L$, $f_T$ ($f_{\perp}$ and $f_{\parallel}$). At BES-III, these
measurements will become available. 
 
As discussed at the beginning of Section~\ref{sec:Charm_decay},  we can obtain the $e^+e^- \rightarrow \DD$ cross sections
by scaling $N_{\DzDzb}$ and $N_{\DpDm}$ by the luminosity at each
energy point. At $E_{cm} = 3773$ MeV, CLEO-c collaboration found peak cross sections
of $\sigma(e^+e^- \rightarrow \DzDzb ) = (3.60 \pm 0.07^{+0.07}_{-0.05})$nb, $\sigma(e^+e^- \rightarrow 
 \DpDm) = (2.79 \pm 0.07^{+0.10}_{-0.04})$ nb, $\sigma(e^+e^- \rightarrow \DD ) = (6.39\pm 0.10^{+0.17}_{-0.08})$ nb, 
 and the ratio of charged $D$ pairs and neutral $D$ pairs production is about
$0.776 \pm 0.024^{+0.014}_{-0.006}$~\cite{cleo-c-cross},
 where the uncertainties are statistical and systematic, respectively. 
The ratio is significantly deviated from one in $\DD$ pair production near threshold.
 It is mainly due to the substantial mass
difference between the charged and neutral $D$ mesons, which
will produce a factor of $\displaystyle \frac{p^3_{+-}}{p^3_{00}} = 0.69$, where
$p_{+-}$ and $p_{00}$ are the momentum of charged and neutral $D$ mesons
in $\psi(3770)$ rest frame. Recently, Voloshin pointed out~\cite{voloshin} that this ratio
should exhibit a prominent variation across the $\psi(3770)$ resonance due
to the interference of the resonance scattering phase with the Coulomb
interaction between the charged $D$ mesons.  The
energy dependent ratio $R$ is expressed by~\cite{voloshin} : 
\begin{equation}
R(E_{cm}) = \frac{\sigma(e^+e^- \rightarrow \DpDm)}{\sigma(e^+e^- \rightarrow
\DzDzb )} = F_c \frac{p^3_{+-}}{p^3_{00}},
\label{eq:coulomb}
\end{equation}
where $F_c$ is the correction factor of the Coulomb interaction, it is 
a function of relative phase of electromagnetic and strong
interactions as described in ref.~\cite{voloshin}. The expected variation of the
ratio in the vicinity of $\psi(3770)$ peak is about a few percent level,
which may be sufficient for a study in the upcoming BES-III experiment.

It is quite well known~\cite{pdg} that the
 cross sections of $e^+e^- \rightarrow D^* \overline{D}^*$, $D^* \overline{D}$, $D_S^{*+} D^{*-}_S$  and  $D_S^{*+} D^{-}_S$ 
in the region just above the threshold of open charm production around 4.0 
GeV display an intricate behavior which is yet to be studied in detail.
This behavior is caused by the successive onset of specific
channels with the $D$ mesons by the strong dynamics in each of these
channels and by the coupling between them~\cite{voloshin_2}. Thus, a
detailed experimental study of this region at BES-III will provide
rich information about the strong dynamics of systems with heavy and light
quarks.    
 
\subsection{$\psi(3770)$ non-$\DD$ Decays}

It is well known that the $\psi(3770)$ decays
most copiously into the OZI-allowed $\DD$ pair owing to the closeness
of the mass threshold. Hadronic or radiative transitions to
lower-lying $c\overline{c}$ states, decay to lepton pairs, or decay
to light hadrons are all available and
predicted~\cite{nora,kuang,rosner}, but their branching fractions are highly suppressed. 
BES-II reported the first signal of non-$\DD$ decays of $\psi(3770)$,  at $\sim 3\sigma$ significance, 
with ${\cal B}(\psi(3770) \rightarrow \pi^+\pi^-J/\psi) = 
(0.34 \pm 0.14\pm 0.09)\%$~\cite{bes-non}.  CLEO-c has
also searched for the non-$\DD$
decays of $\psi(3770)$, including $\pi\pi J/\psi$, $\gamma \chi_{c1}$ and light hadron final
states~\cite{cleo-c-non}. 
The CLEO-c collaboration measured 
cross section for $e^+ e^- \rightarrow \psi(3770) \rightarrow $ hadrons at 
 $E_{cm} = 3773$ MeV to be $(6.38\pm
0.08^{+0.41}_{-0.30})$ nb~\cite{non-dd-cleoc}. The difference
 between this and the $e^+e^- \rightarrow \psi(3770) \rightarrow \DD$ cross section at the same energy is
 found to be $(-0.01 \pm
0.08^{+0.41}_{-0.30})$ nb, which indicates the non-$\DD$ decays would be
smaller than 0.53 nb at 90\% CL, or corresponds to an upper
limit of 8.4\% of the branching fraction at 90\% CL.  BESII
also reported an upper
limit of 30\% of the branching fraction at 90\%
CL~\cite{besii-non-dd}. 

In order to understand the decay dynamics
of non-$\DD$ decays and the line-shape of $\psi(3770)$ resonance, it is
very important to measure the cross sections of inclusive
hadron productions by using scan method in the vicinity of $\psi(3770)$ peak.
At BES-III, the sensitivity to non-$\DD$ measurement is estimated to be less than 1\% level by
using scan method with one year integrated luminosity at different
energy points near $\psi(3770)$ peak.  A detailed MC simulation is
in process.  

The exclusive charmless $\psi(3770)$ decay modes should be searched at
BES-III.  The sensitivity to exclusive charmless $\psi(3770)$ decay modes
at BES-III will be around $10^{-6} - 10^{-7}$ with 20 fb$^{-1}$ data.  
One has to consider the interference between resonances and continuum, and
also the interference between different resonances
near $\psi(3770)$~\cite{mo}.

\section{$D^0-\Dzb$ Mixing, $CP$ Violation and Physics Beyond the Standard Model}
   
With the design luminosity of $10^{33}$cm$^{-2}$s$^{-1}$, BES-III will have
the opportunity to probe for the possible new physics which may enter
up-type-quark decays. It includes searches for charm mixing, $CP$ violation and rare charm
decays. The BES-III charm physics program also includes a variety of
measurements that will improve the determination of $\phi_3/\gamma$ from
$B$-factory experiments.  The total number of charm mesons accumulated
at BES-III will be much smaller than that
at $B$-factories which are about 500 fb$^{-1}$ for each of them.
However, the quantum correlations in the $\psi(3770) \rightarrow
\DD$ system will provide a unique laboratory in which to study
charm~\cite{asner}.

\subsection{$D^0- \Dzb$ Mixing and $CP$ Violation}

$D^0 - \Dzb$ mixing within the SM are highly suppressed due to GIM
mechanism, thus, at BES-III, searches for neutral charm mixing and
$CP$ violation in charm decays may be essential in deciding if
some intriguing signals are actually due to new physics. 

The time evolution of $D^0 -\Dzb$ system, assuming no $CP$ violation
in mixing, is governed by four parameters: $x=\Delta m/\Gamma$ and
$y = \Delta \Gamma/2\Gamma$ which are the mass and width differences of
 $D$ meson mass eigenstates and characterize the mixing matrix, $\delta$
 the relative strong phase between Cabibbo favor (CF) and doubly-Cabibbo
suppressed (DCF) amplitudes and $R_D$ the DCF decay rate relative to the
CF decay rate. The mixing rate $R_M$ is defined as
$\frac{1}{2}(x^2+y^2)$~\cite{asner_2}. Standard Model
based predictions for $x$ and $y$, as well as a variety of
non-Standard Model expectations, span several orders
of magnitude~\cite{asner_2} which is $x$ $\sim$ $y$ $\sim$ $10^{-3}$.
Presently, experimental information about charm
mixing parameters $x$ and $y$ comes from the
time-dependent analyses. The current experimental upper limits
on $x$ and $y$ are on the order of a few times $10^{-2}$. 

At BES-III, time-dependent analyses are not available.
However, one can use the fact that $\DzDzb$ pairs
in $\psi(3770) \rightarrow \DD$ decays have the useful property
that the two mesons are in the $CP$-correlated states~\cite{asner}, namely,
one $D$ state decayed into the final state with definite $CP$ properties
immediately identifies or tags $CP$ properties of the state on the other
side. It provides time-integrated sensitivity to $R_M$ at $10^{-4}$ level
by considering the decay $\psi(3770) \rightarrow (K^-\pi^+)(K^-\pi^+)$
hadronic final state only at BES-III.  We have not estimated the double
semileptonic channel yet, $\psi(3770) \rightarrow (l^\pm (KX))(l^\pm (KX))$. 
Sensitivity to $cos\delta$ is 0.03 for $D^0 \rightarrow
K\pi$ mode.  Recently, a simultaneous determinations of the
mixing, relative strong phase and $R_D$ have been proposed
by using various single tag, double tag and $CP$ tag rates at
CLEO-c~\cite{asner}. The sensitivities on $x$ and $y$ are improved greatly, but
we have not estimated this method at BES-III yet. 

For the direct $CP$ violation,
the SM predictions are as large as 0.1\% for $D^0$ decays, and 
1\% level for $D^+$ and $D_S$ decays~\cite{buccella}. At BES-III, one can also
look at the $CP$ violation by exploiting the quantum coherence
at the $\psi(3770)$. Consider the case where both the $D^0$ and the $\Dzb$
decay into $CP$ eigenstates, then the decays
$\psi(3770) \rightarrow f^i_+ f^i_+$ or  $f^i_- f^i_-$
are forbidden, where $f_+$ ($f_-$) denotes a $CP+$ eigenstate ($CP-$
eigenstate). This is because $CP(f^i_\pm f^i_\pm) = CP(f^i_\pm)CP(f^i_\pm)(-1)^l =
-1$,  while, for the
$l=1$ $\psi(3770)$ state, $CP(\psi(3770)) = +1$. Thus observation of a final state such
as $(K^+K^-)(\pi^+\pi^-)$ constitutes evidence of $CP$
violation. For $(K^+K^-)(\pi^+\pi^-)$ mode, the sensitivity at BES-III
is about 1\% level. Moreover, all pairs of $CP$ eigenstates, where
both eigenstates are even or both are odd, can be summed over for $CP$
violation measurements at BES-III.

\subsection{Dalitz Plot Analyses}

Recent studies of multi-body decays of $D$ mesons provide a direct probe of the final state
interactions by looking at the interference between intermediate state resonances on the Dalitz Plot
(DP). When $D$ mesons decays into three or more daughters,
intermediate resonances dominate the decay rates. These resonances
will cause a non-uniform distribution of events in phase space
on the DP. Since all events on the DP have the
same final states, different resonances at the same location on DP
will interfere. This provides the opportunity to measure both the
amplitudes and phases of the intermediate decay channels, which in
turn allows to deduce their relative branching fractions. These phase
differences can even allow details about very broad resonances to be
extracted by observing their interferences with other intermediate states. 

The most important thing is that recent studies of multi-body decays
of $D$ mesons probe a variety of physics including light
spectroscopy ($\pi\pi$, $K\pi$ and $KK$ S-wave states), searches for $CP$ violation and
 $D^0 -\Dzb$ mixing.  Currently, the decay
$D^0 \rightarrow K_S \pi^+\pi^-$ plays very important role
in the determination of $\phi_3 / \gamma$. Recently BaBar and
Belle~\cite{babar} have reported $\gamma = (70 \pm
31^{+12+14}_{-10-11})^o$ and $\phi_3 = (77^{+17}_{-19} \pm
13 \pm 11)^o$, respectively, where the third error is the systematic error
due to modeling of DP. The precision
of these measurements will eventually be limited by the understanding of the 
$D^0 \rightarrow K_S \pi^+\pi^-$ decays. Although K-matrix description of
the $\pi\pi$ S-wave may yield improved models of the DP and the error on
$\phi_3 / \gamma$ may be decreased from $\pm 10^o$ to a few degrees, it is still a
model-dependent way to extract the angle. At BES-III,
by using the coherence of $\DzDzb$ pairs at
$\psi(3770)$ peak, one can study the CP-tagged and flavor-tagged DP
by doing binned analysis~\cite{bondar}. This method is a model-independent. 
According to the estimation in reference~\cite{bondar}, the proposed
super-$B$ factory~\cite{super-b} with its design integrated luminosity
of $50$ ab$^{-1}$, would allow a measurement of $\phi_3 / \gamma$ with
accuracy below $2^o$. To keep the uncertainty due to $D$ DP decays below
that level, around 10$^{-4}$ $CP$-tagged $D$
decays are needed, corresponding to $\sim 10$ fb$^{-1}$ data which can
be obtained at BEPC-II with two years' luminosity. 
\begin{table}[htbp]
\caption{Current and projected 90\%-CL upper limits on
rare $D^+$ decay modes at BES-III with 20 fb$^{-1}$ data at $\psi(3770)$
peak. We assume the selection efficiencies for all modes are 35\%.}
\label{tab:rare}
\begin{tabular}{@{}llll}
\hline
                & Reference     & Best Upper  &  BES-III  \\
Mode            & Experiment              & limits($10^{-6}$)      & ($\times 10^{-6}$)  \\
\hline
$ \pi^+ e^+e^-$     & CLEO-c~\cite{cleo-c-rare} & 7.4 & 0.03  \\
$ \pi^+ \mu^+\mu^-$ & FOCUS~\cite{focus}  & 8.8 & 0.03  \\
$ \pi^+ \mu^+ e^-$ & E791~\cite{e791}  & 34 & 0.03  \\      
$ \pi^- e^+ e^+$ & CLEO-c~\cite{cleo-c-rare}  & 3.6 & 0.03  \\      
$ \pi^- \mu^+\mu^+$ & FOCUS~\cite{focus}  & 4.8 & 0.03  \\
$ \pi^- \mu^+ e^+$ & E791~\cite{e791}  & 50 & 0.03  \\      
$ K^+ e^+e^- $ & CLEO-c~\cite{cleo-c-rare}  & 6.2 & 0.03  \\    
$ K^+ \mu^+\mu^- $ & FOCUS~\cite{focus}  & 9.2 & 0.03  \\     
$ K^+ \mu^+ e^- $ & E791~\cite{e791}  & 68 & 0.03  \\    
$ K^- e^+ e^+ $ & CLEO-c~\cite{cleo-c-rare}  & 4.5 & 0.03  \\
$ K^- \mu^+ \mu^+ $ & FOCUS~\cite{focus}  & 13 & 0.03  \\
$ K^- \mu^+ e^+ $ & E687~\cite{e687}  & 130 & 0.03  \\
\hline
\end{tabular}
\end{table}
\begin{table}[htbp]
\caption{Current and projected 90\%-CL upper limits on
rare $D^0$ decay modes at BES-III with 20 fb$^{-1}$ data at $\psi(3770)$
peak.}
\label{tab:rare_d0}
\begin{tabular}{@{}llll}
\hline
                & Reference     & Best Upper  &  BES-III  \\
Mode            & Experiment              & limits($10^{-6}$)      & ($\times 10^{-6}$)  \\
\hline
$\gamma \gamma $  & CLEO~\cite{cleo-d0} & 28 & 0.05  \\
$\mu^+\mu^-$ & D0~\cite{d0-coll}  & 2.4 & 0.03  \\
$\mu^+ e^-$ & E791~\cite{e791}  & 8.1 & 0.03  \\      
$e^+ e^-$ & E791~\cite{e791}  & 6.2 & 0.03  \\      
$\pi^0 \mu^+\mu^-$ & E653~\cite{e653}  & 180 & 0.05  \\
$\pi^0 \mu^+ e^+$ & CLEO~\cite{cleo-d01}  & 86 & 0.05  \\      
$\pi^0 e^+e^- $ & CLEO~\cite{cleo-d01}  & 45 & 0.05  \\    
$K_S \mu^+\mu^- $ & E653~\cite{e653}  & 260 & 0.1  \\     
$K_S \mu^+ e^- $ & CLEO~\cite{cleo-d01}  & 100 & 0.1  \\    
$K_S e^+ e^- $ & CLEO~\cite{cleo-d01}  & 110 & 0.1  \\
$\eta \mu^+ \mu^- $ & CLEO~\cite{cleo-d01}  & 530 & 0.1  \\
$\eta \mu^+ e^- $ & CLEO~\cite{cleo-d01}  & 100 & 0.1  \\
$\eta e^+ e^- $ & CLEO~\cite{cleo-d01}  & 110 & 0.1  \\
\hline
\end{tabular}
\end{table}
\subsection{Rare Charm Decays}

Searches for rare-decay processes have played an important role
in the development of the SM. Short-distance flavor-changing
neutral current (FCNC) processes in charm decays are much more highly
suppressed by the GIM mechanism than the corresponding down-type quark decays because
of the large top quark mass. Observation of $D^+$ FCNC
decays $D^+ \rightarrow \pi^+ l^+l^-$ and $D^+ \rightarrow K^+ l^+l^-$ could therefore provide indication of new physics or of
unexpectedly large rates for long-distance SM processes like
$D^+ \rightarrow \pi^+ V$, $V \rightarrow l^+l^-$, with real
or virtual vector meson $V$. Recently, CLEO-c report the branching
fraction of the resonant decay ${\cal BR}(D^+ \rightarrow \pi^+
\phi \rightarrow \pi^+ e^+e^-) = (2.8 \pm 1.9 \pm 0.2 ) \times 10^{-6}$.
The lepton-number-violating (LNV) or lepton-flavor-violating (LFV) decays
$D^+ \rightarrow \pi^- l^+l^+$, $K^- l^+
l^+$ and $\pi^+ \mu^+ e^-$ are forbidden in the SM. Past searches have set upper
limits for the dielectron and dimuon decay modes~\cite{pdg}. 
In Table~\ref{tab:rare} and Table~\ref{tab:rare_d0}, the current
limits and expected sensitivities at BES-III are
summarized for $D^+$ and $D^0$, respectively.  Detailed
description on rare charm decays can be found in references~\cite{Burdman,ian}.
The charm meson radiative decays are also very important to
understand final state interaction which may enhance the decay rates. 
In Ref.~\cite{Burdman,ian}, the decay rates of $D \rightarrow V \gamma$ ($V$ can
be $\phi$, $\omega$, $\rho$ and $K^*$ ) had
been estimated to be $10^{-5} - 10^{-6}$, which can be reached at BES-III. 
\section*{Acknowledgment}
The author would like to thank David Asner for many useful discussions and
comments.

\end{document}